\documentclass[english,prd,floatfix,amsmath,amssymb,11pt,a4paper,nofootinbib]{article}

\usepackage{jheppub}

\usepackage[T1]{fontenc}
\usepackage[latin9]{inputenc}
\usepackage{array}
\providecommand{\tabularnewline}{\\}
\usepackage{babel}

\usepackage{amsmath,amssymb}
\usepackage[margin=10pt,font=small]{caption}
\usepackage{graphicx}
\usepackage{epstopdf}
\usepackage{enumerate}
\usepackage{slashed}
\usepackage{wasysym}
\usepackage{multirow}
\usepackage{array}
\usepackage{mciteplus}
\usepackage{tabularx}
\usepackage{verbatim}
\usepackage{subfig} % must be included after the caption package
\usepackage{hyperref} % has to be last package
\usepackage[all]{hypcap} % references point to top of figures rather than captions
\newcommand{\MTTWO}{\ensuremath{M_{T2}}}
\newcommand{\MTTWOCUT}{\ensuremath{{M_{T2}^{\mathrm{cut}}}}}
\newcommand{\beq}{\begin{equation}}
\newcommand{\eeq}{\end{equation}}
\newcommand{\bea}{\begin{eqnarray}}
\newcommand{\eea}{\end{eqnarray}}
\newcommand{\GeV}{\ensuremath{\mathrm{GeV}}}
\newcommand{\TeV}{\ensuremath{\mathrm{TeV}}}

\newcommand{\met}{\ensuremath{\slashed{p}_T}}
\newcommand{\meff}{\ensuremath{M_{\mathrm{eff}}}}

\newcommand{\inverse}[1]{#1\ensuremath{^{-1}}}
\newcommand{\ttbar}{\ensuremath{t\bar{t}}}

\newcommand{\ignoreSevenTeV}{520}% obtained from requiring fractional statistical uncertainty < 50%
\newcommand{\ignoreFourteenTeV}{780}% obtained from requiring fractional statistical uncertainty < 50%
\newcommand{\lowMTTWOCUTlowlumi}{200}
\newcommand{\highMTTWOCUTlowlumi}{240}

\newcommand{\htdtotal}{57}
\newcommand{\htdCatA}{16}
\newcommand{\htdCatB}{15}
\newcommand{\htdCatC}{16}
\newcommand{\htdCatD}{7}
\newcommand{\htdCatCD}{6}
\newcommand{\htdVeryHard}{Five}
\newcommand{\nrPoints}{1940}
\newcommand{\nrPointsTotal}{1941}

\newcommand{\StablePointsStau}{13}
\newcommand{\StablePointsChiOnePlus}{7}
\newcommand{\StablePointsChiTwoZero}{1}

\newcommand{\definmath}[2] {\def#1{\ifmmode#2\else$#2$\fi}}
\definmath{\pslash}{{\slashed{p}}}
\definmath{\ptslash}{{\slashed{p}_T}}
\definmath{\ptmiss}{{\slashed{\bf p}_T}}
\definmath{\qtmiss}{{\slashed{\bf q}_T}}
\definmath{\pt}{{p_T}}
\definmath{\Pt}{{{\bf p}_T}}
\definmath{\cht}{\tilde{\chi}}
\definmath{\ntlone}{{{\cht^0_1}}}
\definmath{\gluino}{{{\tilde{g}}}}

\onecolumn
\title{Discovery reach for generic supersymmetry at the LHC: \boldmath\MTTWO\ versus missing transverse momentum selections for pMSSM searches}

\author[a]{Benjamin C. Allanach,}
\emailAdd{B.C.Allanach@damtp.cam.ac.uk}
\affiliation[a]{University of Cambridge, DAMTP, Centre for Mathematical Sciences\\Wilberforce Road, Cambridge, CB3 0WA, United Kingdom}

\author[b]{Alan J. Barr,}
\emailAdd{a.barr1@physics.ox.ac.uk}

\author[b]{Alexandru Dafinca}
\emailAdd{a.dafinca1@physics.ox.ac.uk}

\author[b]{and Claire Gwenlan}
\emailAdd{c.gwenlan1@physics.ox.ac.uk}

\affiliation[b]{University of Oxford, Sub-department of Particle Physics\\Denys Wilkinson Building, Keble Road, Oxford, OX1 3RH, United Kingdom}

\abstract{
Different search strategies for supersymmetry have been employed by the LHC general-purpose experiments using early data. As proven by their early results, these strategies are promising, but raise the
question of how well they will generalize for the future. We address
this question by studying two thousand phenomenological minimal
supersymmetric standard model parameter space points that come from
a fit to indirect and cosmological data. We examine the 5$\sigma$
discoverability of the points employing a typical ATLAS-type search
based on missing transverse momentum (MET), a search based on an
optimised \MTTWO\ cut and a combination of the two, taking into
account standard model backgrounds. The discovery reach of the strategies can depend strongly on the systematic uncertainty in the background, subject to the stringency of the cuts and the details of the background simulation. By combining the MET and \MTTWO\ based strategies, with an
integrated luminosity of 1~\inverse{fb} (10~fb$^{-1}$) at 7~TeV,
4-8$\%$ (42$\%$) of the points are discoverable, depending on the
systematic uncertainty on the background. At 14~TeV and with
10~fb$^{-1}$, 96$\%$ of the points are discoverable. While the
majority of points can be discovered by both strategies at
$\sqrt{s}~=~14~\TeV$ and with 1~fb$^{-1}$, there are some that are left undiscovered by a MET search strategy, but which are discovered by the \MTTWO\ strategy, and vice versa, therefore it is essential that one performs both in
parallel. We discuss some of the factors that can make points more
difficult to observe.
}


\arxivnumber{1105.1024}

\begin{document}

\maketitle
\flushbottom

\section{Introduction}\label{Sec:Intro}

The search strategies for supersymmetry at the LHC experiments ATLAS
\cite{Aad:2009wy} and CMS \cite{Bayatian:942733} have been designed
to be as general as possible. At the time of their design they were most extensively tested by the experiments in simulations on 7 and 14 benchmark points, respectively. These points are thought to
be representative for whole extended regions in the minimal
supersymmetric standard model (MSSM) parameter space, but in fact
belong to one particular constrained model called mSUGRA/CMSSM.
Therefore it is possible that these points do not cover all of the
MSSM parameter space satisfactorily. The search strategies (for
example cuts) have been optimized to increase the efficiency of the
searches in these particular cases. It is important to understand to
what extent these searches will be able to discover SUSY in other,
less constrained regions of the MSSM parameter space.

 The searches are based on the observation that in a wide range of SUSY models, squarks and gluinos are expected to cascade decay to jets, leptons or photons. The final state is expected to contain stable, weakly interacting (and therefore invisible) heavy particles, which make a good candidate for dark matter. These will escape detection, giving rise to missing transverse momentum in the detector. SUSY searches at the LHC \cite{Collaboration:2011qk}, \cite{Khachatryan:2011tk} are heavily based on this signature, requiring large missing transverse momentum in order to discriminate standard model processes from supersymmetric events. In our paper, we call such a search a MET-based strategy.

Another strategy examined in this paper is incited by the fact that many well-motivated models of new physics have a $Z_2$ parity (for example $R$-parity in supersymmetry), which predicts that supersymmetric particles must be produced in pairs. Each of the two supersymmetric particles cascade decays within the detector to the lightest parity-odd particle, which is stable and assumed to be weakly interacting. $R$-parity therefore gives rise to very particular event topologies. The \MTTWO\ variable \cite{Lester:1999tx, Barr:2003rg} is a kinematic variable that uses experimentally measurable quantities (the missing transverse momentum being one of them) efficiently, by exploiting the topology of the event, to extract information about the masses of the produced particles. Search strategies using \MTTWO\ as a discriminating variable have been designed \cite{Barr:2009wu} and recently applied by the ATLAS experiment in \cite{Collaboration:2011qk}.

The latest results from the ATLAS \cite{Collaboration:2011qk} and
CMS \cite{Khachatryan:2011tk} experiments have proven that their
search strategies for supersymmetry are promising. However, we would
like to understand how well these generalize to the future. Two recent studies \cite{Conley:2010du, Conley:2011nn} have examined
ATLAS MET based search strategies in a more general context by
simulating signals for $\sim 71$k model points within the
phenomenological MSSM (pMSSM). The landscape of the pMSSM parameter
space has also been explored in a study by AbdusSalam et al.
\cite{AbdusSalam:2009qd}, who have performed a global Bayesian fit
of the pMSSM to current indirect collider and dark matter data. The
parameters of the pMSSM were constrained in the fit, resulting in
inferences on sparticle masses. In the present paper, we take a
sample of points from the fit of reference \cite{AbdusSalam:2009qd}
with a probability proportional to their posterior probability
function. Extending the work of \cite{Conley:2010du, Conley:2011nn}, which do not examine the discovery reach of \MTTWO\ based searches, in this paper we will use the points sampled from reference \cite{AbdusSalam:2009qd} to:
\begin{itemize}
    \item determine how likely not only MET-based searches, but also \MTTWO-based searches are to discover SUSY.
    \item understand the successes and failures of the searches.
    \item optimise the searches for the more general pMSSM scenario.
\end{itemize}
The plan of this report is as follows. First we give a short account
of the properties of the SUSY points examined, sampled by
\cite{AbdusSalam:2009qd}. Then, the MET and \MTTWO\ based search
strategies and the corresponding event selection are explained in
sections \ref{Sec:MET_searches} and \ref{Sec:Mt2}. We briefly present
in section \ref{Sec:Technicalities} the technicalities of simulating
the events for different model points, a basic detector simulation,
a discussion of the Standard Model backgrounds and provide a
definition of \textit{discoverability}.
Section~\ref{Sec:discoverability} shows how the SUSY discoverability
varies with the choice of the \MTTWO\ cut for different centre of
mass energies (7 and 14~TeV) and a variety of integrated
luminosities, in the context of a search strategy based on \MTTWO\
alone and a combined search strategy based on both MET and \MTTWO.
We discuss the most recent ATLAS exclusion limits on sparticle
masses and compare them with the expectations from our study in
section~\ref{Sec:ATLAS_exclusion}. The effect of systematic
uncertainties in the background on discoverability is detailed in
section~\ref{Sec:systematics}. Optimal cuts are proposed in
section~\ref{Sec:Optimal_Cuts}. We discuss what makes a point
difficult to discover in section~\ref{Sec:DifficultPoints}. Finally,
section \ref{Sec:discussion} contrasts our work on the \MTTWO\
search strategy with the analysis from references \cite{Conley:2010du, Conley:2011nn} performed on other ATLAS MET-based search
strategies. We conclude in section~\ref{Sec:conclusions}.

\section{Points investigated}\label{Sec:Points}

A total of \nrPoints\ points were provided by the authors of \cite{AbdusSalam:2009qd}. They are sampled from the posterior probability distribution obtained through a Bayesian fit of electroweak, search and flavour physics data to the pMSSM. The pMSSM contains the most relevant 25 weak-scale MSSM parameters. The priors used in the analysis are flat in mass parameters except for the scalar mass parameters, which are flat in the logarithm. Inferences are made about these points without assuming a restrictive high-scale supersymmetry breaking model.
\begin{table}[ht]
  \begin{center}
    \begin{tabular}{ l | c }
      Parameter & Allowed range\\ \hline
      Scalar masses & $[100~\GeV, 4~\TeV]$\\
      Trilinear scalar coupling & $[-8~\TeV, 8~\TeV]$\\
      Gaugino masses & $[-4~\TeV, 4~\TeV]$\\
      $\tan \beta$ & $[2, 60]$ \\
    \end{tabular}
    \caption{Allowed ranges for the pMSSM parameters. $\tan \beta$ is the ratio of the MSSM Higgs vacuum expectation values.} \label{tb:SUSYpoints_params}
  \end{center}
\end{table}
The ranges of allowed values for some of the parameters are summarized in table \ref{tb:SUSYpoints_params}. Five of the 25~pMSSM parameters varied were associated with the Standard Model: the top mass, the mass of the $Z$ boson, the bottom quark mass and the strong coupling constant. The relic density of dark matter inferred from cosmological measurements proved to be the most important constraint. Various observables involving bottom quarks were used to constrain the model: the branching ratios of $B_S\rightarrow \mu^+ \mu^-$, $B_u \rightarrow \tau \nu$ and $B \rightarrow X_s \gamma$ as well as the $B_s-{\bar B}_s$ mass difference and the $B \rightarrow K^* \gamma$ isospin asymmetry. The electroweak observables used were: the $W$ boson mass, the mass and decay width of the $Z$ boson, the effective weak mixing angle, $Z-$pole asymmetry parameters and the muon anomalous magnetic moment.
Direct pre-LHC searches for sparticles and Higgs bosons were also used.
The 25 dimensional parameter space was successfully scanned with the {\tt MultiNest} algorithm~\cite{Feroz:2007kg,Feroz:2008xx}, achieving a statistically convergent fit. The resulting fit was far from being robust, since there were not enough data to constrain so many parameters, and this manifests itself in the high degree of prior dependence in the results. In the present paper, we are not concerned with the issue of fit robustness: we merely want to use a subset of the points that fit current data well in order to investigate the properties of proposed search strategies in a general context, as explained in Section~\ref{Sec:Intro}.

\begin{figure*}
\begin{centering}
\includegraphics[angle=0, width=1\textwidth]{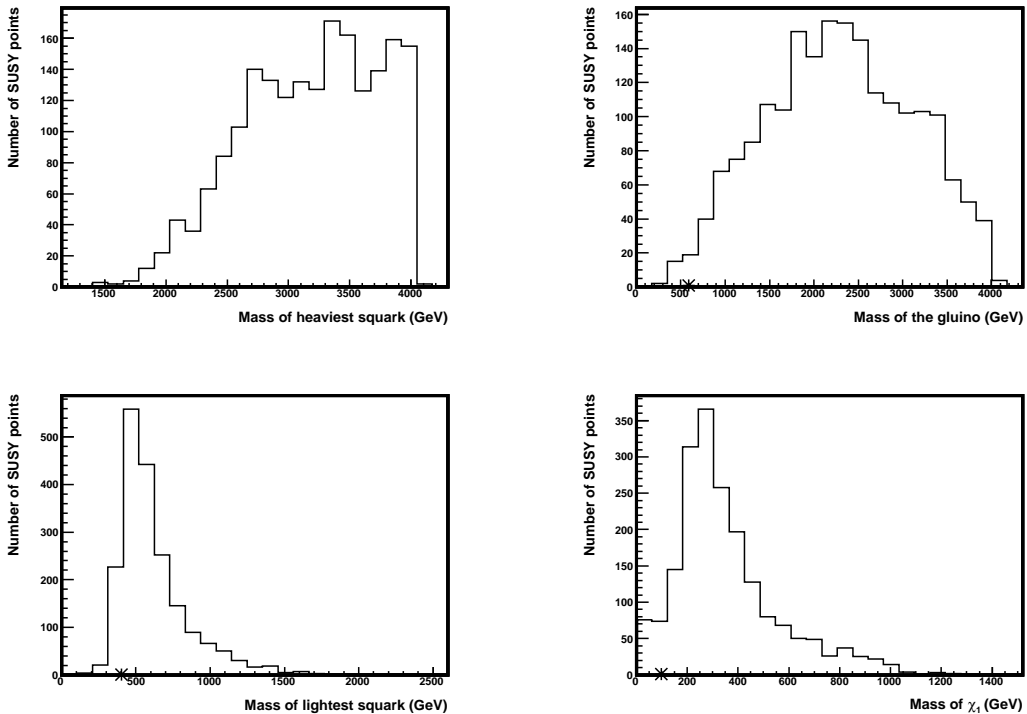}
\caption{A flavour of the mass spectra of the \nrPointsTotal\ SUSY points investigated. The * denotes the SPS1a point. A * is missing from the first histogram since the heaviest squark of the SPS1a point ($m = 586$~\GeV) is much lighter than the corresponding squark in any of the other points examined.\label{Fig:SUSYspectra}}
\par\end{centering}
\end{figure*}

Together with the SPS1a point \cite{Allanach:2002nj}, these points form the pool of scenarios studied in this report. A flavour of their mass spectra is given in figure \ref{Fig:SUSYspectra}. In the sampling from the pMSSM, the physical masses of the sparticles was constrained to lay below 4~TeV. In contrast, all seven mSUGRA/CMSSM benchmark points used by the ATLAS collaboration in \cite{ATLAS:1278474} have a gluino mass below 1~TeV. Notice that the SPS1a point, which is used widely as a benchmark point, is also in the low mass end of the spectra and is now excluded by the ATLAS experiment~\cite{Collaboration:2011qk}.

\section{MET based search strategy}\label{Sec:MET_searches}

The final state of supersymmetric events is expected to contain stable, weakly interacting (and therefore invisible) heavy particles. These will escape detection, giving rise to missing transverse momentum in the detector. In this paper we study one possible discovery channel, in which we require 4 or more jets and 0 leptons in the final state. The 0 lepton requirement suppresses the (semi)leptonic decays of $W$, $Z$ and \ttbar, whereas the large number of jets suppresses the QCD background. The concrete selection for this channel follows the recipe laid out in \cite{Aad:2009wy}. We require:

\begin{itemize}
\item at least four jets with transverse momentum $\pt~>~50$~\GeV, at least one jet must have $\pt~>~100$~\GeV
\item missing transverse momentum $\met~>~100$~\GeV
\item no isolated leptons with $\pt~>~20$~\GeV
\item the smallest azimuthal separation between the jet direction and the missing transverse momentum, $\Delta\phi\left(\mathrm{jet}_i,\ptmiss\right)~>~0.2$ for the first three jets
\item effective mass $\meff~>~800$~\GeV
\item $\met~>~0.2~\meff$
\item transverse sphericity $S_T~>~0.2$
\end{itemize}

The missing transverse momentum \ptmiss\ is calculated as the negative
sum of the transverse momenta of all visible objects within detector
acceptance:
\begin{equation}\label{eq:met_def}
\ptmiss \equiv - \sum_{i}\Pt^{(i)}
\end{equation}
where we define the acceptance as having
pseudorapidity $\left|\eta\right| < 5$ and only take into account objects with
$\pt~>~0.5$~\GeV.\footnote{$\eta =
-\ln\left[\tan\left(\frac{\theta}{2}\right)\right]$, where $\theta$
is the angle between the particle 3-momentum and the beam axis.}

The effective mass \meff\ is defined as the scalar sum of transverse
momenta of the most energetic objects in the event - in this case
four jets, together with the missing transverse momentum:
\begin{equation}\label{eq:meff_def}
\meff \equiv \sum_{i=1}^{n=4}\left|\Pt^{(i)}\right| + \met
\end{equation}

The transverse sphericity $S_T$ is defined as:
\begin{equation}
S_T \equiv \frac{2\lambda_2}{\left(\lambda_1+\lambda_2\right)}
\end{equation}
with $\lambda_1$ and $\lambda_2$ being the eigenvalues of the $2
\times 2$ sphericity tensor $S_{ij} = \sum_{k}
\left(p_{T}^{(k)}\right)_{i} \left(p_{T}^{(k)}\right)_{j}$. The
tensor is computed using all jets with $\left| \eta \right|~<~2.5$
and $\pt~>~20$~\GeV.

A simplified version of this strategy, suitable for early data, was
employed by ATLAS for three of the four signal regions defined in
reference~\cite{Collaboration:2011qk}. These follow the same
philosophy as reference \cite{Aad:2009wy}, but require a reduced jet
multiplicity and slightly looser cuts to suit the limited
statistics. The fourth signal region of
reference~\cite{Collaboration:2011qk} employed a different selection
variable described in the following section.

\section{\MTTWO\ based search strategy}\label{Sec:Mt2}

The \MTTWO\ variable is a generalization of the transverse mass to the case of decays of pairs of particles \cite{Lester:1999tx, Barr:2003rg}. Each of the two decays will contribute to the final state with one visible and one invisible object. The transverse momentum $\Pt^{(1)}$ and $\Pt^{(2)}$ of each of the visible objects can be measured in the detector. However, only the \textit{total} transverse momentum of the invisible objects can be measured as missing transverse momentum \ptmiss. Nevertheless, for each of the decays labeled by $i=1,2$ one can compute the transverse mass $m_{T}^{(i)}$ under the assumption of a massless invisible object:
\begin{equation}\label{eq:mtdef}
{m_{T}^{(i)}}^2 (\Pt^{(i)}, \qtmiss^{(i)}) \equiv \\
     2\left|\Pt^{(i)}\right|\left|\qtmiss^{(i)}\right|-2\Pt^{(i)}\cdot\qtmiss^{(i)}\
\end{equation}
where $\qtmiss^{(i)}$ is a guess for the true, unknown missing transverse momentum $\ptmiss^{(i)}$. The variable \MTTWO\ is defined by:\begin{equation}\label{eq:mttwodef}
\MTTWO (\Pt^{(1)}, \Pt^{(2)},\ptmiss) \equiv \\
     \min_{\sum{\qtmiss} = \ptmiss} \left\{ \max\left( m_T^{(1)}, m_T^{(2)} \right) \right\} \
\end{equation}
The minimization takes place over all values of the two undetectable particles' possible missing transverse momenta $\qtmiss^{(1,2)}$ consistent with the constraint $\sum{\qtmiss} = \ptmiss$.

The variable \MTTWO\ gives the greatest lower bound on the parent mass for a given event that is consistent with the observed momenta and hypothesised masses \cite{Cheng:2008hk}. When plotted over all events with the same parent particles, one expects an endpoint to appear. For example, for \ttbar\ production, we expect \MTTWO\ to have an endpoint at the mass of the top quark. While the endpoint will be smeared by effects such as detector resolution, simulations \cite{Barr:2009wu} show that the vast majority of standard model background events have $\MTTWO < m_t$. For SUSY events however, where the actors are far heavier particles, one can expect \MTTWO\ to have larger values. This allows good discrimination between signal and background for the region $\MTTWO > m_{\mathrm{top}}$.\footnote{We note that for fairly degenerate mass spectra, where the splitting between sparticles is less than the top mass, good discrimination may not be possible.} Motivated by this observation, the use of the \MTTWO\ variable as a complement to the standard ATLAS search strategies has been advocated in \cite{Barr:2009wu} and is now being actively pursued by the ATLAS collaboration, for example in \cite{ATLAS-CONF-2010-065,
Collaboration:2011qk}.

\begin{figure*}
\begin{centering}
\includegraphics[width=1\columnwidth]{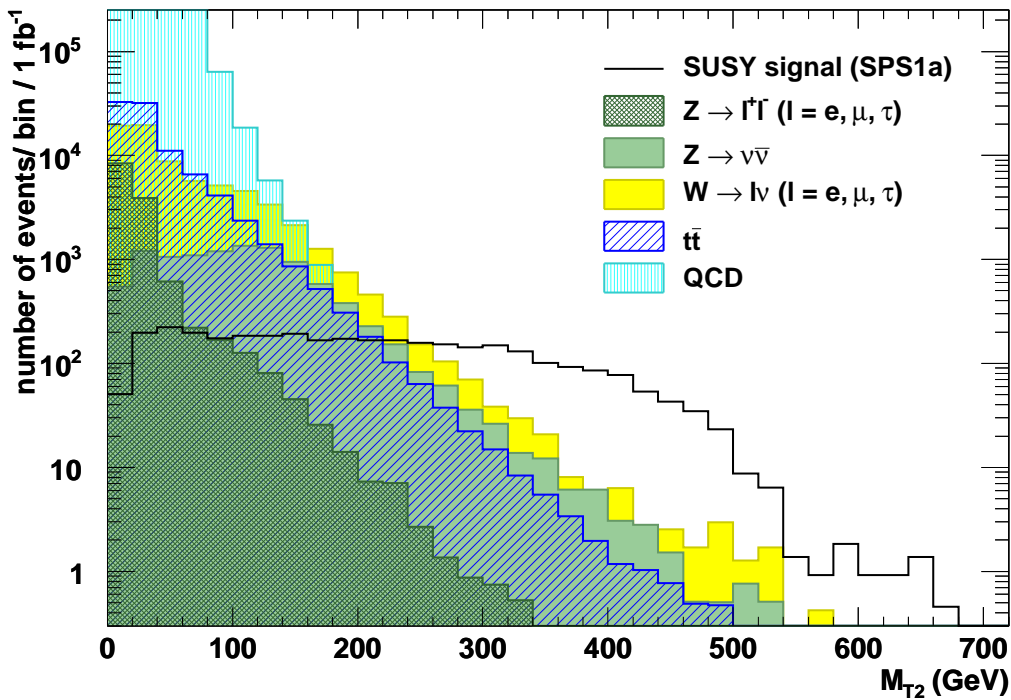}
\caption{Distribution of \MTTWO\ for events with two or more jets
with $p_{T}~>~50$~\GeV\ (and no other cuts) for an integrated
luminosity of 1~\inverse{fb} at $\sqrt{s}~=~7$~\TeV.\label{Fig:SPS1a_background}}
\par\end{centering}
\end{figure*}

Typical \MTTWO\ spectra for simulated backgrounds and for the SPS1a SUSY scenario are displayed in figure~\ref{Fig:SPS1a_background}, reproducing results of~\cite{Barr:2009wu}. These were obtained as described in section~\ref{Sec:Technicalities}. The figure shows that for a \MTTWO\ cut $\gtrsim 600$~GeV placed too high, not only the background, but also the signal is suppressed. For values of the \MTTWO\ cut too low, the background swamps the signal. Our goal is to strike a balance between these two extremes. Based on the SPS1a point one can find the range of values for a cut on \MTTWO\ which will render the SPS1a model discoverable, i.e. which 1) provides a sufficient number of signal events after selection and 2) provides a small enough \mbox{$p$ -- value} for the background-only hypothesis. The meaning of ``sufficient number'' and ``small enough'' will be defined in the next section. SPS1a is only one possible scenario of how nature could look, so when determining the optimal value for the cut on \MTTWO, we perform the same analysis on many other scenarios.

ATLAS has implemented in one of the four signal regions in reference \cite{Collaboration:2011qk} an \MTTWO\ search strategy by requiring no leptons, a jet multiplicity $\geq$ two, with the leading jet $\pt\ > 120$~\GeV, second jet $\pt > 40$~\GeV, $\met > 100$~\GeV, $\Delta\phi \left(\mathrm{jet},\ptmiss\right) > 0.4$ and $\MTTWO > 300$~\GeV. The \MTTWO\ variable is calculated using the two highest-\pt\ jets. The \met\ requirement makes an effective trigger, while the $\Delta \phi$ requirement reduces the QCD background.

In this paper we take a simplified version of this analysis, with the intention of examining the effect of a cut on \MTTWO\ alone. We only require two or more jets with $\pt > 50$~\GeV, compute \MTTWO\ using the two highest-\pt\ jets and then cut at different values of \MTTWO. We call the value of \MTTWO\ at which the cut is set \MTTWOCUT. No restriction is set on the number of leptons.
\section{Technical details}\label{Sec:Technicalities}

The following operations have been performed in this analysis:

\begin{itemize}
\item The spectrum information for each of the \nrPoints\ pMSSM points sampled by the authors of \cite{AbdusSalam:2009qd} and for the SPS1a point was passed to {\tt Herwig++~2.4.2}~\cite{Bahr:2008pv,Gieseke:2011na} via the SUSY Les Houches Accord~\cite{Skands:2003cj}. For each point we simulated $10\,000$ inclusive SUSY events. A sample of background events was produced, containing $10^7$ events of each \ttbar, $W \rightarrow l\nu+\mathrm{jets}$, $Z \rightarrow l^{+}l^{-}+\mathrm{jets}$, $Z \rightarrow \nu\bar{\nu}+\mathrm{jets}$. For QCD we produced $10^7$ events for each slice of \pt\ of hard scatter, where the slices lie between 17, 35, 70, 140, 280, 560, 1120~GeV, with the last slice starting at  2240~GeV. This is to ensure we have sufficient simulated events at high \pt.  The contribution from diboson+jets is expected to be very small according to \cite{Aad:2009wy}. For each of the SUSY scenarios and the background, we further performed the following:
\item For each event, we clustered hadrons with fiducial pseudorapidity ($\left|\eta\right|~<~5$) and momentum ($p_T~>~0.5~\GeV$) into jets using the fastjet \cite{Cacciari:2005hq} implementation of the \mbox{anti-$k_T$} algorithm \cite{Cacciari:2008gp}. The $E$ combination scheme with $R~=~0.4$ and ${p_T}^{\mathrm{min}}~=~10~\GeV$ was used.
\item We simulated the effect of the detector in a simplified way by smearing the energy and momenta of the jets using a Gaussian probability function for the majority of $\left( 1-\epsilon \right)$ events. The width of the Gaussian depends on the energy of the jet such that:
\beq
\sigma\left(E\right)/E_j=\left(0.36/\sqrt{E_j\left[\GeV\right]}\right)\oplus
0.1 \eeq
where $E_j$ is the unsmeared jet energy. This resolution is typical
of general-purpose LHC detectors \cite{Aad:2009wy, Bayatian:942733}.
The distribution also has a low energy tail to account for the
possibility of badly mismeasured jets, therefore the remaining
fraction $\epsilon = 1\%$ of the jets is smeared with a probability
density: \beq \log P(r) = \left\{ \begin{array}{ll} c_1 r + c_2\ &
\mathrm{for} ~ (0.2<r<0.8) \\ 0 &
\mathrm{elsewhere}\end{array}\right. \eeq
where $r \equiv E/E_j$ is
the ratio of the smeared jet energy to the true jet energy. The
constant $c_1$ is chosen such that the resulting smearing function
of a jet of given energy agrees with full simulation
results as in~\cite{Aad:2009wy} and $c_2$ is a normalization constant.

The missing transverse momentum is calculated from the negative vector sum of the jet momenta (after smearing) and other isolated particles present.
\end{itemize}

In addition, for the MET and \MTTWO\ search strategy individually,
we determined the number of events passing the cuts defined in
sections~\ref{Sec:MET_searches} and~\ref{Sec:Mt2}. For the \MTTWO\
based strategy we allowed for a variable cut on \MTTWO. We then
compared the number of signal events with the number of background
events to determine whether the point in question would be
discoverable at a given luminosity. For a SUSY point to be
\textit{discoverable}, we require on one hand at least 10 SUSY
events to pass the cuts. On the other hand, the signal should be
well above background, i.e. the \mbox{$p$ -- value} for the
background-only hypothesis should be low enough. We will convert the
\mbox{$p$ -- value} into an equivalent significance $Z$, defined such
that a $Z$ standard deviation upward fluctuation of a Gaussian
random variable would have an upper tail area equal to $p$. That is,
$Z = \Phi^{-1}\left(1-p\right)$, with $\Phi$ being the cumulative
distribution of the Standard Gaussian. We then require for a
discovery that $Z>5$. We can include the systematic uncertainty in
the background by using the methods from reference
\cite{Cousins:2008zz}.

Having decided for each SUSY point individually for which range of \MTTWOCUT\ it would be discoverable and whether the MET search strategy would discover the point, we would like to know the answers to the following questions:
\begin{itemize}
\item What is the overall optimal value for \MTTWOCUT?
\item How is the optimal value of \MTTWOCUT\ affected as we include systematic uncertainties in the background?
\item How is the optimal value of \MTTWOCUT\ affected as the experiments accumulate more luminosity?
\item How does the optimized \MTTWO\ search compare to the MET search when taking into account the above?
\end{itemize}
We will address these questions in the following sections.

\section{$\int\mathcal{L}dt$-dependent discoverability}\label{Sec:discoverability}

\begin{figure*}[ht]
\centering
\subfloat[$\sqrt{s} = $ 7~TeV. Systematic uncertainties in the
background have been
neglected.\label{Fig:7TeVdiscoverability_0.001}]{
\includegraphics[angle = 0, width=0.475\textwidth]{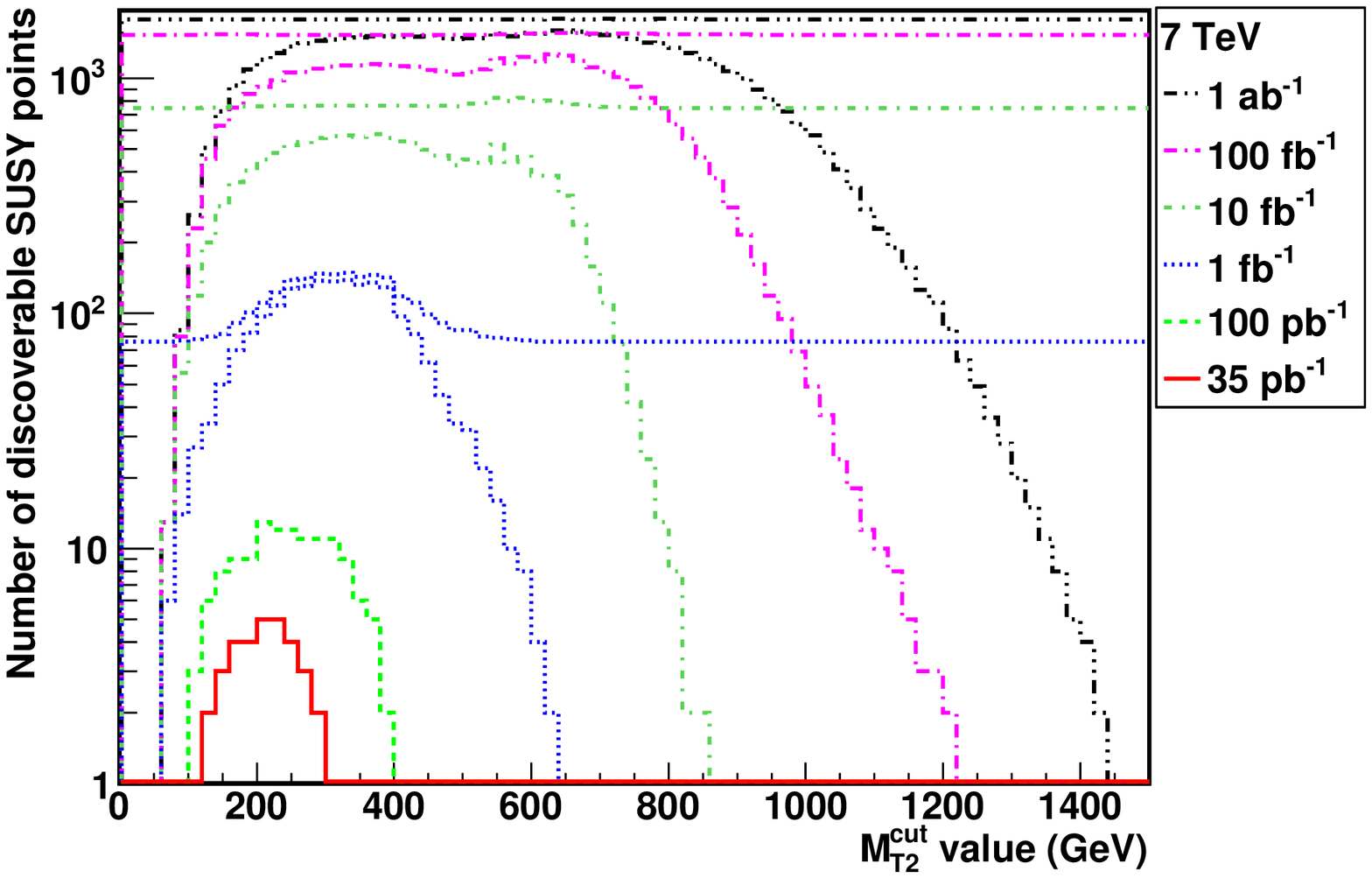}} \subfloat[$\sqrt{s} = 14 $~TeV. Systematic uncertainties in the
background have been
neglected.\label{Fig:14TeVdiscoverability_0.001}]{\includegraphics[angle=0,
width=0.475\textwidth]{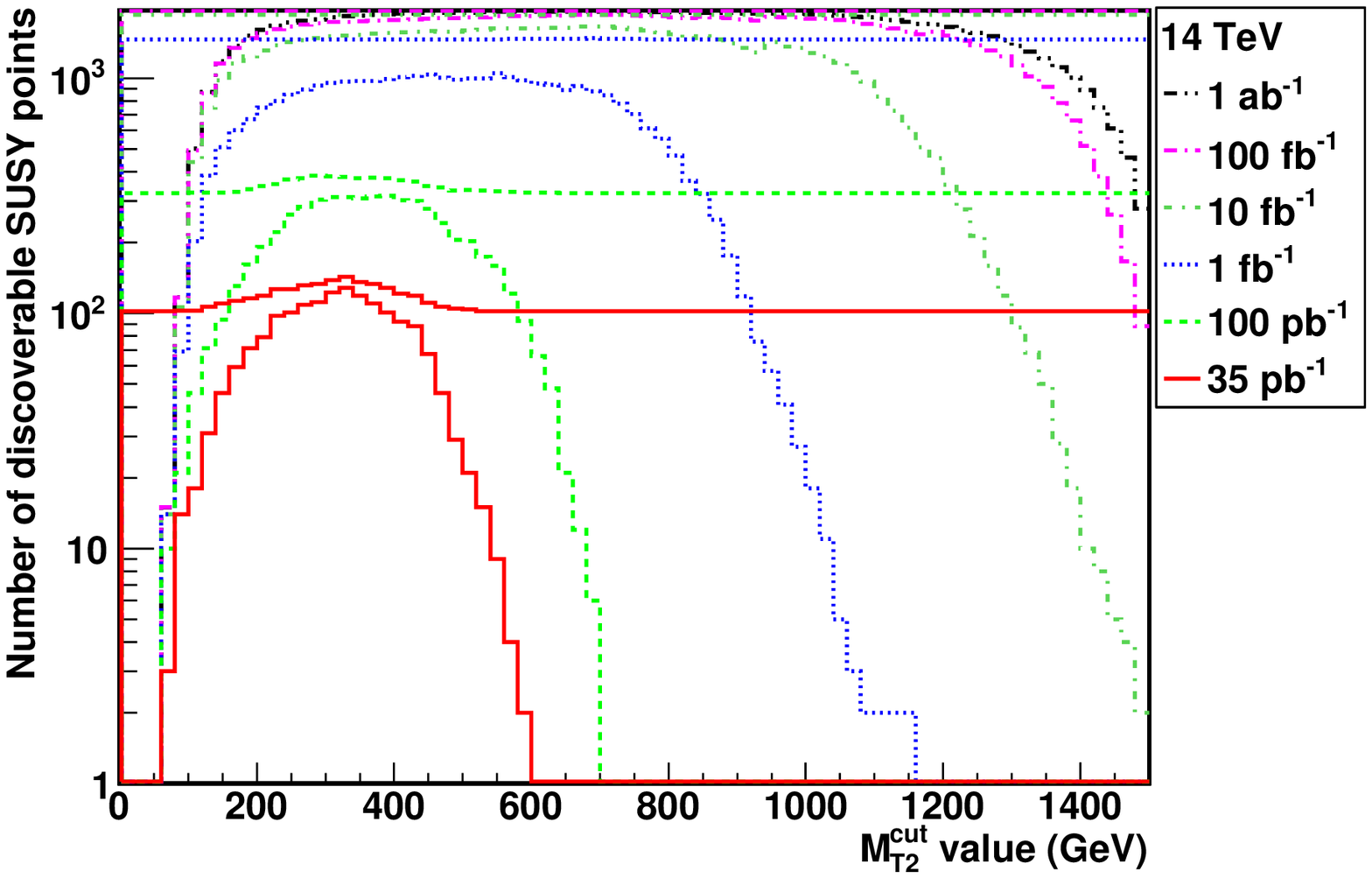}}
\\
\subfloat[$\sqrt{s} = 7$~TeV. The systematic uncertainty in the
background was assumed to be 50\%. Notice no points can be
discovered with
35~\inverse{pb}\label{Fig:7TeVdiscoverability_0.5}]{\includegraphics[angle=0,
width=0.475\textwidth]{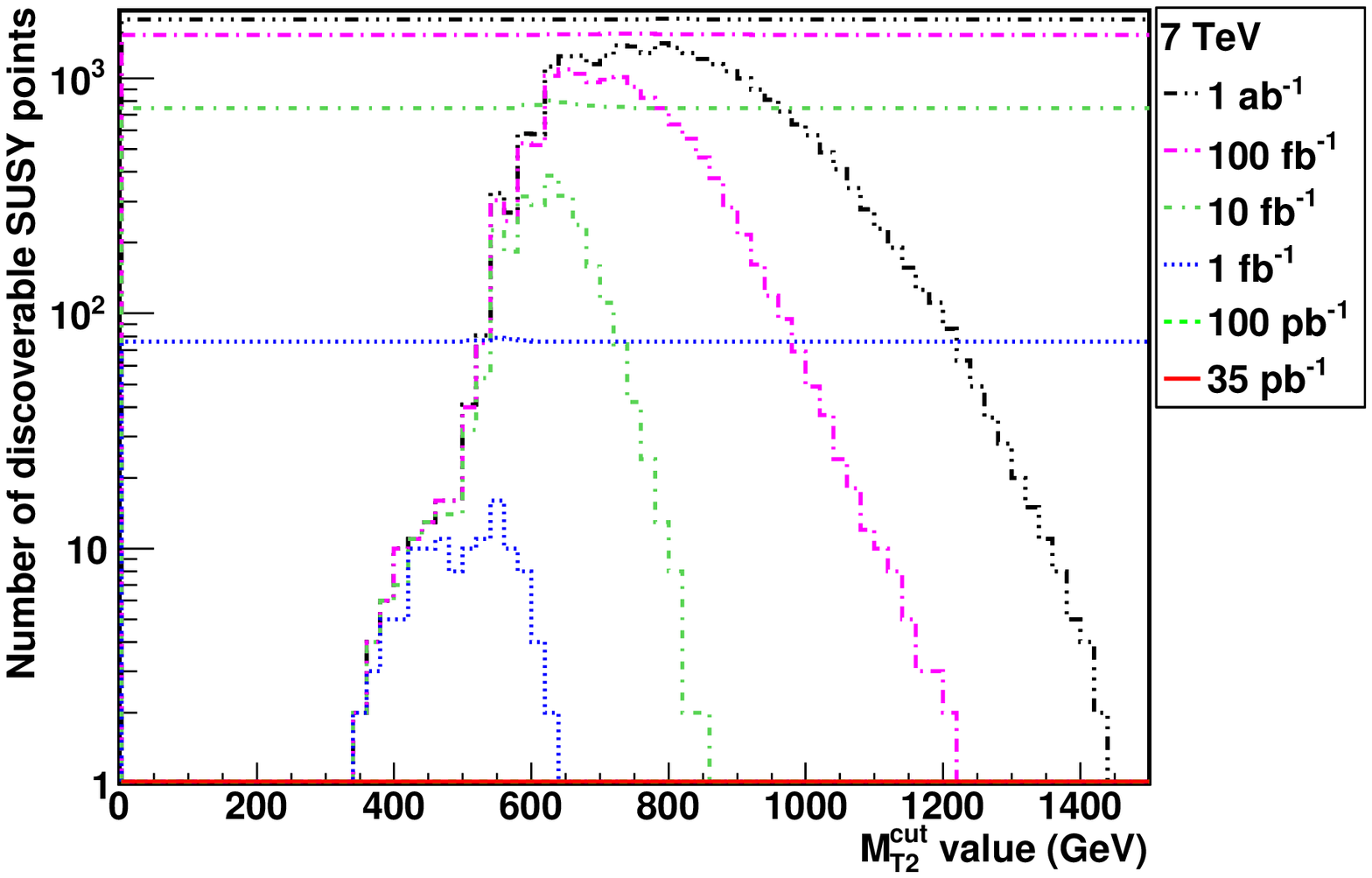}}
\subfloat[$\sqrt{s} = $ 14~TeV. The systematic uncertainty in the
background was assumed to be
50\%.\label{Fig:14TeVdiscoverability_0.5}]{\includegraphics[angle=0,
width=0.475\textwidth]{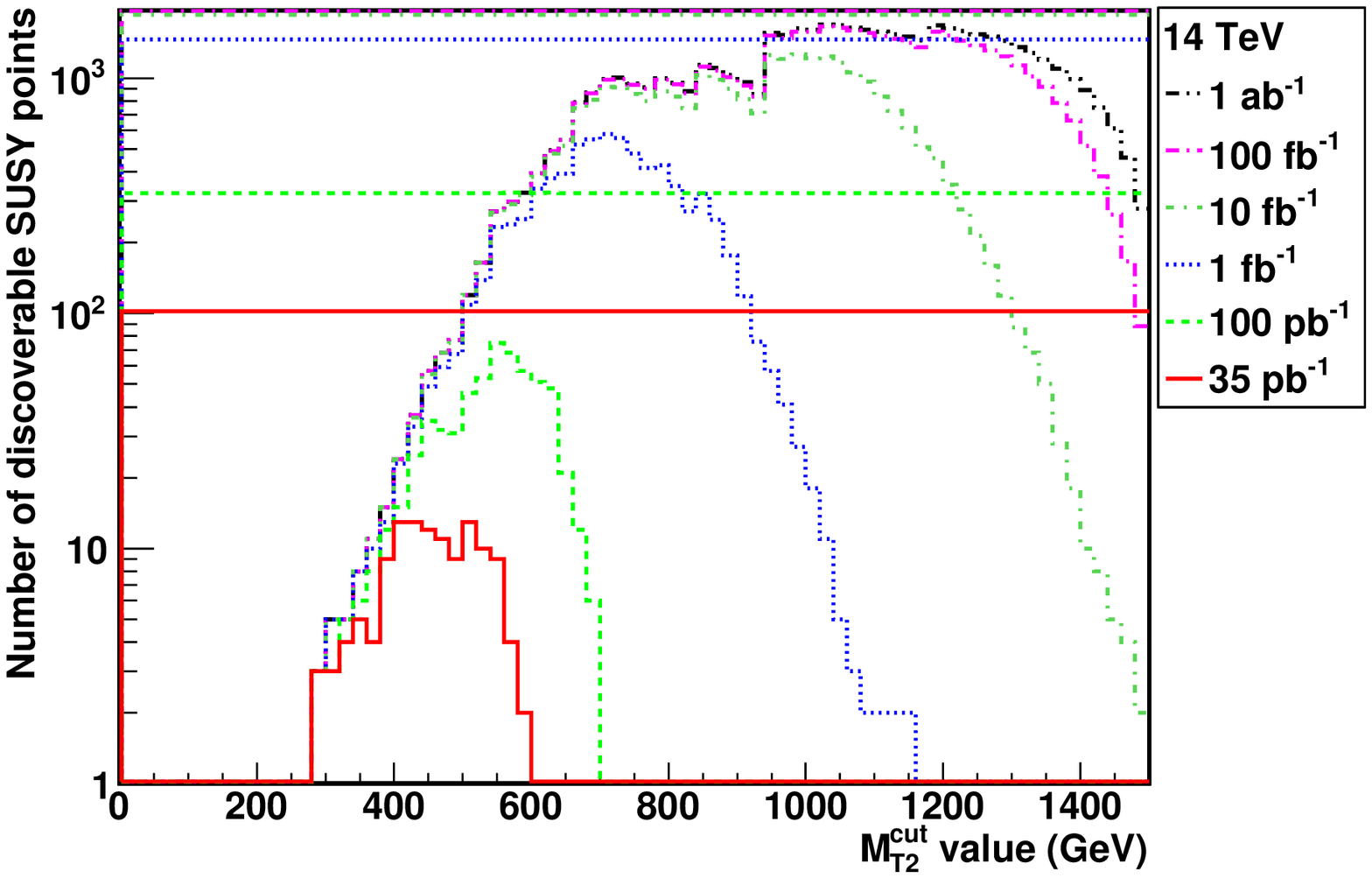}}
\caption{Number of SUSY points (out of \nrPointsTotal) discoverable
at $\sqrt{s} = $ 7 or 14~TeV versus \MTTWOCUT\ (bins of 20~GeV) for
the \MTTWO\ based strategy (lower line of same style) and the
combined strategy based on \MTTWO\ and MET (upper line of same
style). For $\sqrt{s} = $ 7 (14)~TeV, predictions in the region
above \ignoreSevenTeV\ (\ignoreFourteenTeV)~GeV might not be
accurate due to the limited statistics in the background.}
\end{figure*}

At the time of writing, the LHC is scheduled to run at
$\sqrt{s}=7~\TeV$, where it is planned to accumulate a total of a
few \inverse{fb}. Figure \ref{Fig:7TeVdiscoverability_0.001} shows
the number of points discoverable with different luminosities at
7~TeV as a function of \MTTWOCUT. For each integrated luminosity,
two lines of the same style were drawn. The lower one shows the
number of points discoverable with the \MTTWO\ based search strategy
alone. The upper of the two lines with the same style shows
discoverability for a combination of the \MTTWO\ and MET based
strategies, where a point is considered to have been discovered if
it was discovered by \textit{either} the \MTTWO\ or the MET-based
strategy. The flat sections of the upper lines give the number of
points discoverable with the MET based search strategy alone. If no
points were discovered with the MET based strategy, the \MTTWO\ only
and combined strategy lines overlap. Only the histograms
corresponding to an integrated luminosity of up to 1~\inverse{fb}
have been or will be experimentally explored with a center-of-mass
energy of 7~TeV.

The same study can be applied for the LHC at its design center of
mass energy of 14~TeV. Figure \ref{Fig:14TeVdiscoverability_0.001}
shows the discoverability in this scenario. Due to the increased
center of mass energy, high discoverability is achieved with far
lower integrated luminosity.

Notice in figure \ref{Fig:SPS1a_background} that our background
sample is statistics limited at the high \MTTWO\ end, giving rise to
a ragged, discontinuous spectrum. This translates in figures
\ref{Fig:7TeVdiscoverability_0.001}-\ref{Fig:14TeVdiscoverability_0.5}
into possibly fake peaks of discoverability at large \MTTWO\ values.
Therefore, one should view critically predictions of discoverability
for $\sqrt{s} = 7~(14)~\TeV$ for values of \MTTWOCUT\ above
\ignoreSevenTeV~GeV (\ignoreFourteenTeV~GeV).

Once the discoverability profile and an optimal \MTTWOCUT\ value is
found through simulations for a given integrated luminosity, it is
not necessarily obvious if the same profile or optimal \MTTWOCUT\
value will apply for different luminosities. In the limit of large
numbers of events and no systematic uncertainties in the background,
the significance varies as $\propto \sqrt{\mathcal{L}}$. This implies that as more luminosity is accumulated, discoverability
will increase at low \MTTWOCUT\ values, where high background is a
problem for the significance. At the high \MTTWOCUT\ end,
discoverability is not necessarily limited by the significance as
much as by the fact that not enough signal events will have been
recorded. Higher integrated luminosity increases discoverability for
very high \MTTWOCUT\ values as well, since the number of signal
events is proportional to luminosity. Considering these two effects
together, we expect the "peak" of the discoverability distribution
with respect to \MTTWOCUT\ to broaden and to shift to higher \MTTWO\
values. This explains the very flat maximum observed at high
luminosity in figures \ref{Fig:7TeVdiscoverability_0.001} and
\ref{Fig:14TeVdiscoverability_0.001}. One can choose a low \MTTWO\
value which optimizes the search at low luminosity - the flat
maximum will ensure that the same value will provide a nearly
optimal strategy at higher luminosity as well.

We have studied discoverability by incrementing the luminosity by a
factor of 10 each time. However, this does not buy nearly the same
factor in discoverability. This is understandable since on one hand
$Z \propto \sqrt{\mathcal{L}}$ at the optimal value of \MTTWOCUT. On
the other hand, if $\sqrt{s} < 2 m_{\mathrm{LSP}}$, where
$m_{\mathrm{LSP}}$ is the mass of the lightest supersymmetric
particle, then sparticles are simply not produced, regardless of the
luminosity.

From the corresponding discoverability plots it becomes apparent
that, in the absence of uncertainties in the background, the
optimized \MTTWO\ based strategy is better suited than the MET based
strategy for integrated luminosities up to several inverse
femtobarns, after which the MET based strategy performs better.

In an attempt to explain this, we found that the MET based search is
too stringent and does not allow any background events at all and
only very few signal events through. Therefore there are plenty of
SUSY points which have fewer than 10 signal events passing the cuts
of the MET based strategy. For these points, the discoverability is
limited by the {\it number} of events, which grows linearly with the
luminosity. On the other hand, the \MTTWO\ based strategy operating
at the value of \MTTWOCUT\ giving maximal discoverability is {\it
significance} limited and $Z$ grows $\propto \sqrt{\mathcal{L}}$.

At low luminosity, not enough signal events pass cuts for the MET
based strategy to be efficient. With increasing luminosity, the MET
based strategy starts to dominate due to the linear growth of the
number of events as opposed to the $\propto \sqrt{\mathcal{L}}$ growth of significance for the \MTTWO\ based strategy. At high
luminosity, of more than a few inverse femtobarns, this effect is so
evident that the combination of the \MTTWO\ and MET based strategies only discovers very few points more than the MET based strategy.

\begin{figure}
\begin{centering}
\includegraphics[width=1\textwidth]{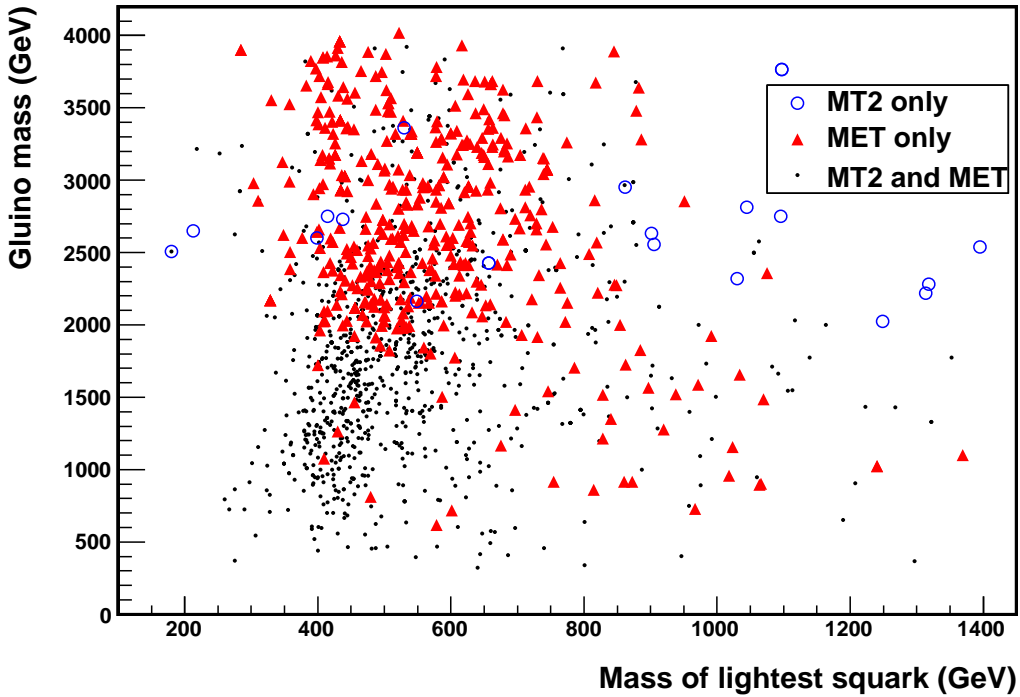}
\caption{Points in the ($m_{\tilde{g}}~-~m_{\mathrm{lightest}~\tilde{q}}$) - plane, discovered by either \MTTWO\ alone, or MET alone, or by both MET and \MTTWO, assuming $\sqrt{s}~=~14~\TeV$,  a luminosity of 1~\inverse{fb} and a fixed systematic uncertainty in the background of 10\%. The \MTTWO\ search operates at its optimum, which for this luminosity and systematic uncertainties is at \MTTWOCUT = 660~GeV.\label{Fig:who_discovers}}
\par\end{centering}
\end{figure}

Figure~\ref{Fig:who_discovers} makes clear that while both the \MTTWO\ and MET based strategies are good at discovering a large sub-space of the pMSSM, at high luminosities above 1~\inverse{fb} the MET based strategy explores the region of heavy gluino masses, $m_{\tilde{g}}~\gtrsim~2~\TeV$, more effectively. Whilst the two strategies have a large overlap of points they discover, there still are regions in the $\left(m_{\tilde{g}} - m_{\mathrm{lightest}~\tilde{q}}\right)$ plane which are only explored by one or the other strategy. There are a variety of reasons for this, which include the inability to set a single cut threshold on \MTTWO\ which is optimal for all points and the relative sensitivity of each strategy to different jet multiplicities, jet $p_T$ and decay topologies. As an example of such a reason, we note that the \MTTWO\ based strategy was designed for the topology of pair-producing SUSY particles, whose direct decays are assumed to produce the hardest two jets in the event. These are the jets that ought to enter the calculation of \MTTWO. However, we found it is possible for one or both of the two hardest jets in certain events to be initial or final state radiation. These events will be more likely to fail the \MTTWO\ than the MET cuts. Similarly, points favoring event topologies where the hardest two jets are daughters of SUSY particles will be discovered with the \MTTWO\ based strategy more easily.
Since the the two strategies are predominantly accepting events of different topologies, a combined strategy tends to be more powerful than the individual strategies.

We conclude this section by noting that so far the LHC SUSY searches
\cite{Collaboration:2011qk, Khachatryan:2011tk} have not been
strongly affected by multiple in-time $pp$ collisions. In the
future, due to the increase in the instantaneous luminosity, the
effects of such pile-up will have to be investigated in dedicated
studies, to include the modified response of the detector.

\section{Exclusion limits from ATLAS}\label{Sec:ATLAS_exclusion}

Turning the problem around, we ask: could one have expected the LHC to exclude SUSY at 95\% confidence level (had it been one of the points in this study), with the integrated luminosity it achieved by the end of 2010, i.e. $\sim 35$~\inverse{pb} and ignoring systematic uncertainties? The most stringent constraints up to now come from the ATLAS collaboration, which excludes in reference~\cite{Collaboration:2011qk} at the 95\% confidence level (CL) points with gluino or squark masses up to 500~GeV in simplified models containing only squarks of the first two generations, a gluino octet and a massless neutralino.

For an \MTTWOCUT\ value between \lowMTTWOCUTlowlumi\ and
\highMTTWOCUTlowlumi~GeV (which is our optimal value at the given
luminosity) the combined strategy from this paper (after an
integrated luminosity of 35~\inverse{pb}) would have provided
evidence for discovery with \mbox{$p$ -- value~$<$~5\%} for 54 points,
including the SPS1a point. These points should have been discovered
or excluded at 95\% CL by the LHC experiments in references
\cite{Collaboration:2011qk,Khachatryan:2011tk}. Indeed, these 54
points have squark and gluino masses that fall within the exclusion
limits of the simplified model of the ATLAS
analysis~\cite{Collaboration:2011qk}. The simplified model assumed
$m_{\tilde{\chi_1}} = 0$, which is the value for which discovery
would be most straightforward, and contain no other light
supersymmetric particles. Whilst it is true that our `discoverable'
model points all have
$(m_{\mathrm{lightest}~\tilde{q}},m_{\tilde{g}})$ values which lie
within the exclusion region of the ATLAS analysis, the inverse is
not true; there are many points in the sample examined in this paper
with {\em squark and gluino masses} that lie within the ATLAS
simplified-model exclusion region which would {\em not} be
discovered with 35~\inverse{pb} -- for a variety of different
reasons. Some characteristic sparticle masses of the five points
which need least luminosity for exclusion are shown in
table~\ref{tb:35Ipb_discovered}. The characteristics of some of the
points that are found to be most difficult to discover with
10~\inverse{fb} at $\sqrt{s}~=~14~\TeV$ are discussed in
section~\ref{Sec:DifficultPoints}. The points studied therein do not
fall within the ATLAS exclusion region, but the reasons for which
they are difficult to discover also apply to the elusive points
lying within the ATLAS exclusion region.

\begin{center}
\begin{table*}[ht]
\begin{tabular}{ r | c | c | c | r }
 Point &$m_{\tilde{g}}$ (\GeV) & \parbox[t]{4cm}{Mass of lightest squark from the first two generations (\GeV)} & \parbox[t]{4cm}{Mass of lightest squark from the third generation (\GeV)} & $m_{\chi_1} (\GeV)$ \\ \hline
  n65 & 520 & 395 ($\tilde{u}_L, \tilde{u}_R, \tilde{c}_L,\tilde{c}_R$) & 848 ($\tilde{b}_1$) & 229\\
  n188 & 560 & 430 ($\tilde{d}_R, \tilde{u}_L, \tilde{s}_R, \tilde{c}_L$) & 437 ($\tilde{b}_1$) & 198\\
  SPS1a & 604 & 548 ($\tilde{d}_R, \tilde{u}_R, \tilde{s}_R, \tilde{c}_R$) & 401 ($\tilde{t}_1$) & 97\\
  p29 & 597 & 406 ($\tilde{u}_L, \tilde{c}_L$) & 1955 ($\tilde{t}_1$) & 183\\
  p780 & 546 & 398 ($\tilde{u}_L, \tilde{c}_L$) & 306 ($\tilde{b}_1$) & 199\\
\end{tabular}
\caption{SUSY points that would have been excluded or discovered at
$5\sigma$ with $\int \mathcal{L} dt = 35$~\inverse{pb} at $\sqrt{s}
= 7~\TeV$ with a combination of the \MTTWO\ and MET based search
strategies as described in this paper. In two of the columns we show
in parentheses the lightest squark flavour belonging to the first two or the third generation.
For the case of the first two squark generations, we group together
several squarks with near-degenerate masses if their mass difference
is less than 1~GeV.} \label{tb:35Ipb_discovered}
\end{table*}
\end{center}
\section{Including systematic uncertainties}\label{Sec:systematics}
Unavoidably, there will be both theoretical and experimental uncertainties in the background. We model the uncertainties in the background by assuming a conservative systematic uncertainty of 50\%, which is comparable to that reported for signal regions B and D in the latest ATLAS SUSY search \cite{Collaboration:2011qk}.

For the study of the effect of systematic uncertainties we assume
the background to be a Poisson process with mean $\hat{b}$, where
$\hat{b}$ itself is a Gaussian-distributed quantity with a standard
deviation of $\mathrm{max}\left\{50\%\times\hat{b},1\ \mbox{ event}
\right\}$. For each SUSY point we calculate in bins of \MTTWOCUT\
the \mbox{$p$ -- value} for the background-only hypothesis, which again
can be translated into a significance as described in reference
\cite{Cousins:2008zz}. With the altered prescription for calculating
the significance, the discoverability is changed to the one
displayed in figures \ref{Fig:7TeVdiscoverability_0.5} and
\ref{Fig:14TeVdiscoverability_0.5}. The intermediate regime of a
systematic uncertainty in the background of 10\% is also considered
in this study. The effect of including the systematic uncertainties
in the background is to push up the optimal value of \MTTWOCUT.

Assuming a constant fraction of 50\% systematic error in
the background, regardless of the value of \MTTWOCUT, is a rather
na\"{i}ve assumption. However, the detailed determination of the
uncertainties in the backgrounds will require both complete detector
simulation and supporting measurements from the LHC data, and so can
only be performed by the experimental collaborations themselves. Our
assumptions on the uncertainties should not affect the qualitative
discussion we present in this paper, however.

\section{Optimal \MTTWOCUT}\label{Sec:Optimal_Cuts}

We will consider the optimal value of the \MTTWO\ cut to be the one which maximizes the number of discoverable points according to the two conditions outlined in section~\ref{Sec:Technicalities}. These conditions take into account both statistical and systematic uncertainties. For the optimum determination we have calculated the number of discoverable points in bins of the \MTTWOCUT\ in sections~\ref{Sec:discoverability} and \ref{Sec:systematics} and found the bin with the most entries.

There are two caveats linked to our determination of the optimum,
neither of which is directly relevant for the experimental
measurements. Firstly, our optimal \MTTWOCUT\ is based on the
limited number of events that we have simulated in the tail of the
\MTTWO\ distribution for both signal and background. Secondly, the
overall systematic uncertainty will vary across the space of the
selection variables, whereas we have assumed it to be constant. What
will be done in practice by the experiments will be to determine
most of the backgrounds and their uncertainties from other
measurements of LHC data.

Table~\ref{tb:MTTWO_MET} summarizes the results for the optimal
\MTTWOCUT\ value in the cases of no systematic uncertainty in the
background, or an assumed 10\% or 50\% systematic uncertainty. The
general trend is that the optimal value of \MTTWOCUT\ goes up with
increasing luminosity, as expected from the discussion in section
\ref{Sec:discoverability}. In figure
\ref{Fig:14TeVdiscoverability_0.001} we observe that the maximum of
discoverability is very flat in \MTTWOCUT, such that choosing an \MTTWOCUT\ value
which maximizes discoverability at say 100~\inverse{pb} will be
approximately optimal for higher luminosities as well. Considering
the above, the following values look reasonable for the choice of
the initial \MTTWOCUT\ value for low luminosities, in the context of
a combination of the \MTTWO\ and MET based search strategies: for
running at 7~TeV, 320~GeV; for running at 14~TeV, 400~GeV.

We also notice that with 50\% uncertainty in the background, the
\MTTWO\ based strategy barely adds any discoverability to the MET
results.

\begin{table*}[ht]

%%%%%%%%%%%%%%%%%%%%%%%%%%%%%%%%%%%%%%%%%%%%%%%

\begin{tabular}{|>{\centering}p{0.15\textwidth}|>{\centering}p{0.15\textwidth}||>{\centering}p{0.15\textwidth}|>{\centering}p{0.08\textwidth}|>{\centering}p{0.08\textwidth}||>{\centering}p{0.08\textwidth}|>{\centering}p{0.08\textwidth}|>{\centering}p{0.08\textwidth}|>{\centering}p{0.08\textwidth}|}

\hline
\multicolumn{2}{|c||}{Energy} & \multicolumn{3}{c||}{7 TeV} & \multicolumn{3}{c|}{14 TeV}\tabularnewline
\hline
\multicolumn{2}{|c||}{Background uncertainty $\sigma_{\hat{b}} / \hat{b} = $} & 0\% & 10\% & 50\% & 0\% & 10\% & 50\% \tabularnewline
\hline \hline

\multirow{3}{*}{$\mathcal{L} = $ 35 pb$^{-1}$} & \MTTWO & 0.3\% & 0.2\% & 0.0\% & 6.6\% & 4.6\% & 0.7\% \tabularnewline
\cline{2-8}
& MET & 0.0\% & 0.0\% & 0.0\% & 5.3\%  & 5.3\% & 5.3\% \tabularnewline
\cline{2-8}
& combined & 0.3\% & 0.2\% & 0.0\% & 7.4\%  & 6.2\% & 5.3\% \tabularnewline
\hline \hline

\multirow{3}{*}{$\mathcal{L} = $ 100 pb$^{-1}$} & \MTTWO & 0.7\% & 0.5\% & 0.1\% & 16.3\% & 12.5\% & 3.9\% \tabularnewline
\cline{2-8}
& MET & 0.0\% & 0.0\% & 0.0\% & 16.7\%  & 16.7\% & 16.7\% \tabularnewline
\cline{2-8}
& combined & 0.7\% & 0.5\% & 0.1\% & 19.9\%  & 17.6\% & 16.8\% \tabularnewline
\hline \hline

\multirow{3}{*}{$\mathcal{L} = $ 1 fb$^{-1}$} & \MTTWO & 7.2\% & 4.1\% & 0.8\% & 54.3\% & 47.6\% & 29.7\% \tabularnewline
\cline{2-8}
& MET & 3.9\% & 3.9\% & 3.9\% & 75.2\%  & 75.2\% & 75.2\% \tabularnewline
\cline{2-8}
& combined & 7.7\% & 5.4\% & 4.1\% & 76.2\%  & 76.2\% & 75.4\% \tabularnewline
\hline \hline

\multirow{3}{*}{$\mathcal{L} = $ 10 fb$^{-1}$} & \MTTWO & 29.9\% & 26.1\% & 19.8\% & 85.8\% & 81.6\% & 65.2\% \tabularnewline
\cline{2-8}
& MET & 38.4\% & 38.4\% & 38.4\% & 95.9\%  & 95.9\% & 95.9\% \tabularnewline
\cline{2-8}
& combined & 42.5\% & 42.5\% & 41.6\% & 96.1\%  & 96.0\% & 96.0\% \tabularnewline
\hline \hline

\multirow{3}{*}{$\mathcal{L} = $ 100 fb$^{-1}$} & \MTTWO & (65.0\%) & (64.6\%) & (56.2\%) & 96.2\% & 92.7\% & 85.4\% \tabularnewline
\cline{2-8}
& MET & (78.8\%) & (78.8\%) & (78.8\%) & 99.5\%  & 99.5\% & 99.5\% \tabularnewline
\cline{2-8}
& combined & (80.3\%) & (80.3\%) & (80.0\%) & 99.6\%  & 99.6\% & 99.6\% \tabularnewline
\hline \hline

\multirow{3}{*}{$\mathcal{L} = $ 1 ab$^{-1}$} & \MTTWO & (82.6\%) & (80.3\%) & (72.6\%) & 99.0\% & 95.9\% & 87.2\% \tabularnewline
\cline{2-8}
& MET & (91.6\%) & (91.6\%) & (91.6\%) & 99.8\%  & 99.8\% & 99.8\% \tabularnewline
\cline{2-8}
& combined & (92.2\%) & (92.2\%) & (92.1\%) & 99.9\%  & 99.9\% & 99.9\% \tabularnewline
\hline

\end{tabular}

\caption{Percentage of discoverable points (out of \nrPointsTotal)
for the optimised \MTTWO\ based, the MET based and the combined
search strategies, different luminosities and
$\sqrt{s}~=~7,~14~\TeV$. The systematic uncertainty in the
background is taken to be a fixed percentage, $f~=~$ 0\%, 10\% or
50\% of the expected number of background events $\hat{b}$, but at
least one event, i.e. $\sigma_{\hat{b}}~=~\mathrm{max} \left\{
f\times\hat{b} \mbox{ events} , 1 \mbox{ event}\right\}$. The
entries in brackets show the fractions of discoverable points for
luminosities which we do not expect to be recorded at those
energies. The number of points discovered with our MET based
strategy does not depend on the systematic uncertainty in the
background due to none of the simulated background events passing
the stringent cuts of this analysis. A more careful analysis
\cite{Conley:2010du, Conley:2011nn} of the expected backgrounds
shows a strong dependence of discoverability on the size of the
systematic uncertainty.}

\label{tb:MTTWO_MET}
\end{table*}

\section{Difficult points}\label{Sec:DifficultPoints}

Some difficult SUSY points evade detection unless they are tackled with a high energy LHC with high luminosity. Why is this the case? There are in principle two types of reasons:

\begin{itemize}
  \item The sparticle masses are very large with a consequently low SUSY production cross-section;
  \item The mass spectrum and branching ratios lead to decay topologies / kinematics with low values of the variables which the search strategies cut on.
\end{itemize}

\begin{figure}
\begin{centering}
\includegraphics[angle=0, width=1\textwidth]{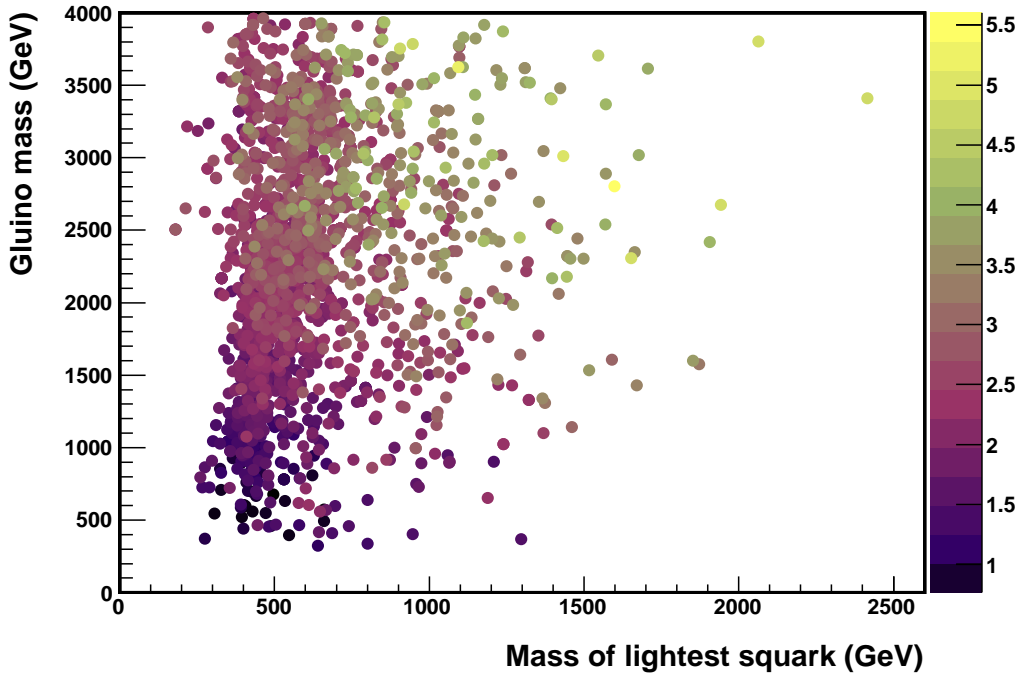}\caption{$\log_{10}$[luminosity (\inverse{pb)} needed for discovery] in the ($m_{\tilde{g}},m_{\mathrm{lightest}~\tilde{q}}$) plane at 14~TeV with a combination of the MET based and optimised \MTTWO\ based strategies. The \MTTWO\ based strategy was optimized for an integrated luminosity of 1~\inverse{fb}. Systematic uncertainties in the background have been neglected.\label{Fig:lumiNeededAt14TeV_MIN}}
\par\end{centering}
\end{figure}

\begin{figure}
\begin{centering}
\includegraphics[angle=0, width=1\textwidth]{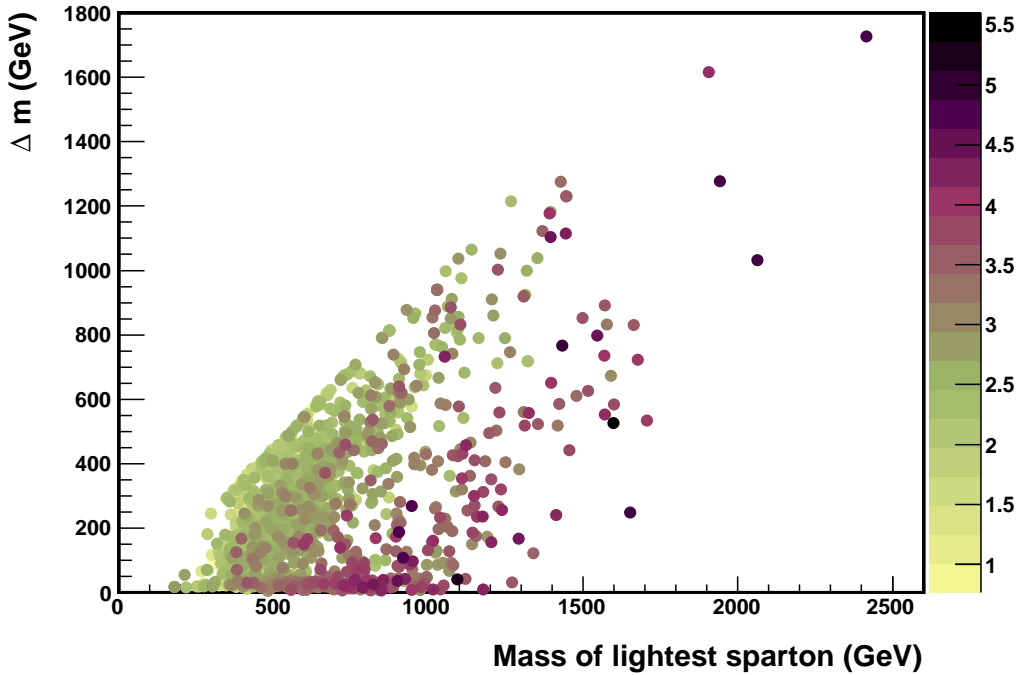}\caption{$\log_{10}$[luminosity (\inverse{pb)} needed for discovery] with the combined optimal \MTTWO\ and MET based strategy at $\sqrt{s}~=~14~\TeV$ in the $(\Delta m, m_{\mathrm{lightest\ sparton}})$ plane. $\Delta m$ is the mass difference between the lightest sparton and the LSP. The \MTTWO\ based strategy was optimized for an integrated luminosity of 1~\inverse{fb}. Systematic uncertainties in the background have been neglected.\label{Fig:lumiVSlsp_Mt2_14TeV}}
\par\end{centering}
\end{figure}

Reasons of the first type need to be tackled with higher
center-of-mass energies and higher luminosities. Reasons of the
second type need to be tackled by improving the search strategies.
Figure \ref{Fig:lumiNeededAt14TeV_MIN} sheds light over which of the
two types of reasons dominate discoverability for the SUSY points
investigated. It shows points in the ($m_{\tilde{g}}$,
$m_{\mathrm{lightest\: }\tilde{q}}$) plane, color-coded to describe
the luminosity needed for discovery in the scenario of
$\sqrt{s}~=~14~\TeV$ with a combination of the MET based and the
optimized \MTTWO\ based search strategy and ignoring systematic
uncertainties in the background. The value \MTTWOCUT~=~320~GeV
optimizes the cut ``in the long run'', i.e. for 1~\inverse{fb} and
above. The points which are hard to discover predominantly populate
high-mass regions of the ($m_{\tilde{g}}$, $m_{\mathrm{lightest
}~\tilde{q}}$) plane. The transition from black (easily discoverable
with $\sim 1$~\inverse{pb}) in the bottom left corner to
yellow\footnote{light gray in a b/w printout} (hard to discover,
need $> 10$~\inverse{fb}) at the extremities does not seem to be
single-valued and figure~\ref{Fig:lumiVSlsp_Mt2_14TeV} examines this
further. It shows the luminosity needed for discovery in the
($\Delta m, m_{\mathrm{lightest\ sparton}}$) plane, with $\Delta m$
being the difference between the mass of the lightest sparton and
that of the LSP. Figure~\ref{Fig:lumiVSlsp_Mt2_14TeV} shows that the
heavier the sparticles, the harder it is to discover the point the
sparticles belong to. However, for the same mass value for the
lightest sparton on the abscissa, some points are still much harder
to discover than others. The variation of luminosity with $\Delta m$
suggests that another important factor for discoverability is the
mass degeneracy of the lightest sparton and the LSP.

\begin{figure*}[ht]
\subfloat[An example of a point where all squarks and the gluino are
very heavy. Gaugino production dominates. (\htdCatA\ such points
identified).]{
\includegraphics[angle=0, width=0.5\textwidth]{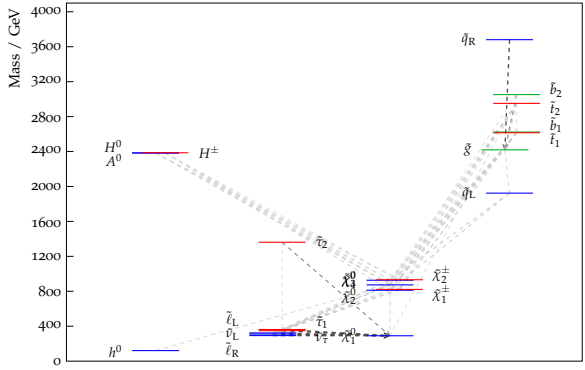}
\label{Fig:spectrum_v_heavy_squarks} } \subfloat[An example of a
point where all sparticles are very heavy (\htdCatB\ such points
identified).]{
\includegraphics[angle=0, width=0.5\textwidth]{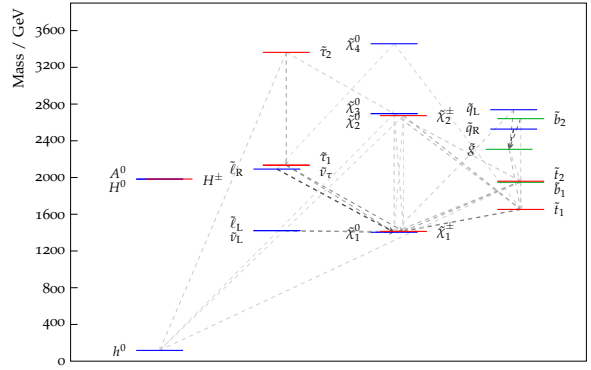}
\label{Fig:heavy_everything} } 
\\
\subfloat[An example of a point with
nearly mass-degenerate squark of the first two generations and LSP
(\htdCatC\ points identified).]{
\includegraphics[angle = 0, width=0.5\textwidth]{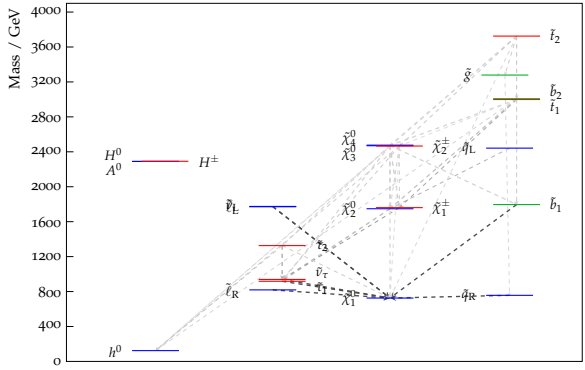}
\label{Fig:spectrum_degenerate_qR_chi} } \subfloat[An example of a
point with nearly mass-degenerate squark of the third generation and
LSP (\htdCatD\ points identified).]{
\includegraphics[angle=0, width=0.5\textwidth]{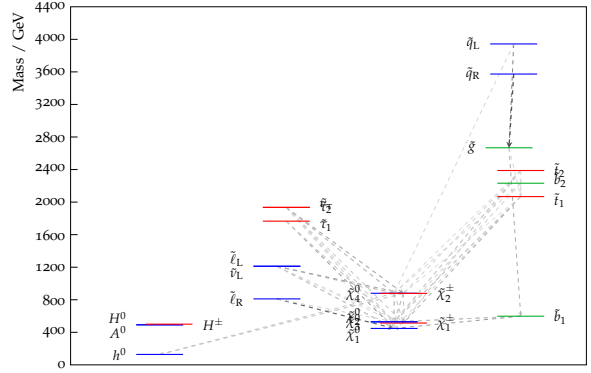}
\label{Fig:spectrum_degenerate_bI_chi} } \caption{Some hard to
discover points. The dotted lines show possible decay modes. Only
branching ratios of above 5\% have been drawn. The darker shades
correspond to higher branching ratios.}
\label{Fig:hard_points_spectra}
\end{figure*}

A few representative spectra of hard-to-discover points are shown in
figure \ref{Fig:hard_points_spectra}, as plotted with
\texttt{PySLHA} \cite{Buckley:PySLHA}. We identified \htdtotal\
points which require more than 10~\inverse{fb} at $\sqrt{s} =
14~\TeV$ to be discovered. Several categories can be distinguished:

1) Points which due to high sparton masses have a small sparton production cross-section of less than 10~fb, while the overall SUSY cross-section is above 10~fb due to slepton and gaugino production. These points cannot be easily discovered at the LHC with any of the ATLAS MET search strategies unless dedicated studies \cite{Lytken:2003ed, Chen:2010ek} are employed. These points might be easier to discover at a future linear collider. A representative mass spectrum of the \htdCatA\ points identified in this category is shown in figure \ref{Fig:spectrum_v_heavy_squarks}.

2) Points which due to the high sparticle masses have a small overall production cross-section of less than 10~fb. These points might prove difficult to discover at a future linear collider as well. A representative mass spectrum of the \htdCatB\ points that fall in this category is shown in figure \ref{Fig:heavy_everything}. 

3) Points which are not cross-section limited, i.e. $\sigma~>~10$~fb.

The above three categories are mutually exclusive. However, the
points can also be categorised depending on the mass degeneracies of
the lightest sparton and the LSP, and we show such spectra in
figures \ref{Fig:spectrum_degenerate_qR_chi},
\ref{Fig:spectrum_degenerate_bI_chi}. Any other squarks and the
gluino are far more massive and their production is therefore
suppressed. Even if the production cross-section for the nearly
mass-degenerate squarks is large enough, above 10~fb, the mass
degeneracy leads to a decay topology which only gives soft jets,
making the signal difficult to distinguish from background. We
identified in our sample \htdCatC\ points which have a lightest
squark of the first two generations nearly mass-degenerate with the
LSP and \htdCatD\ points which have a $\tilde{b}$ or $\tilde{t}$
squark nearly mass-degenerate with the LSP. \htdCatCD\ points fall
in both categories, having both the lightest squark of the first two
generations and the lightest squark of the third generation nearly
mass-degenerate with the LSP. Points with mass near-degenerate particles could be dealt with by looking at events where the produced sparticle pair is boosted by the emission of hard-QCD jets from initial or final state radiation \cite{Alves:2011sq,Alwall:2008ve}.

\htdVeryHard\ hard-to-discover points are both cross-section limited
and exhibit large mass degeneracies, which will make them
particularly hard to tackle at the LHC.

\section{Discussion}\label{Sec:discussion}

Recent papers by Conley et al. \cite{Conley:2010du, Conley:2011nn}
describe similar studies to the present one, applying several other
ATLAS MET based search strategies on a set of $\sim$71k model points
in pMSSM. The similarities and differences between these studies and
ours are discussed in this section.

In the scenarios studied by Conley et al., an additional particle
outside the MSSM is implicitly assumed to make up part or all of the
dark matter relic density. This has the consequence of a
preponderance of wino dominated lightest supersymmetric particles,
leading to quasi-mass degenerate lightest charginos and lightest
neutralinos. Reference \cite{Conley:2011nn} did not perform a global
fit, but instead scanned for points which were compatible with each
experimental 95\% CL. Depending on the scan prior used, the
sparticle masses were constrained to be less than 3~TeV or 1~TeV.
Each point was given equal weight in their results.

In comparison to our work, in the MET based strategies employed by
Conley et al., the lepton multiplicity required is varied. However,
it is suggested both in their work and in ATLAS studies
\cite{Aad:2009wy, Collaboration:2011qk} that the 0 lepton channel is
the most robust of the MET-based analyses. The backgrounds in
references \cite{Conley:2010du, Conley:2011nn} were produced by the
ATLAS SUSY group with state-of-the-art Monte Carlo event generators
and the full ATLAS detector simulation. The signal was generated
using a fast detector simulation.

For different scenarios with systematic uncertainty in the
background of 50\% or 20\%, Conley et al. calculate how many of the
generated SUSY points are discoverable using any of the ATLAS 11 MET
analyses, requiring the same criteria for discoverability as the
ones in the present paper. Conley et. al found an optimal cut on the
value of the \meff\ variable employed in MET type analyses, while we
did the same for the \MTTWO\ variable. It is difficult to perform a
direct analysis between our results and those by Conley et al.,
since their paper starts out from different assumptions. We note for
example that setting the sparticle mass upper range to 1 or
3~TeV would have further increased the fraction of points discovered
in our study.

Reference \cite{Conley:2011nn} found that, at $\sqrt{s}~=~7$~\TeV\
and after 1~\inverse{fb}, MET cuts would discover between 50-90\% of
the models considered, depending on the systematic error assumed on
the background. Here, we found a similar level of dependence of
\MTTWO-based search discoverability on the systematic error assumed.
References \cite{Conley:2010du, Conley:2011nn} did not consider the
\MTTWO\ based search strategy.

Conley et al. also study the effect of a detector-stable charged
LSP, whereas the LSP in the points we investigated was
$\tilde{\chi}_1^{0}$ throughout. We did find though
\StablePointsChiOnePlus\ points with pseudo-stable $\tilde{\chi}_1^{+}$, \StablePointsChiTwoZero\ point with pseudo-stable $\tilde{\chi}_2^{0}$ and \StablePointsStau\ points with pseudo-stable $\tilde{\tau}_1$.\footnote{We consider a particle to be ``pseudo-stable'' if it has a lifetime larger than 100 picoseconds or equivalently its decay length is longer than 3 cm.} These points exhibit
extreme mass degeneracies between the pseudo-stable,
next-to-lightest supersymmetric particle and the LSP. The mass
difference is a few hundred MeV and the decay is strongly
phase-space suppressed. The pseudo-stable particles have decay
lengths from a few centimeters to several meters. Whilst the
simulation that we performed becomes invalid for these points since
it includes no information on the detector geometry, the high
production cross-section for these points combined with a search for
long-lived charged particles would make such points detectable by
other means \cite{Aad:2009wy}.

The analysis of the discovery reach of a search strategy based on
\MTTWO\ is unique to our paper. Also, in our study we have taken
into consideration a bigger range of sparticle masses, allowing for
much higher hierarchies in the points studied. The limited range of
sparticle masses restricts the type of hard-to-discover points found
by references \cite{Conley:2010du, Conley:2011nn}. Both our study
and references \cite{Conley:2010du, Conley:2011nn} investigated the
effects of varying the systematic uncertainty in the background and
the luminosity, but Conley et al. did not explore the discovery
potential at $\sqrt{s}~=~14$~\TeV\ for luminosities higher than
1~\inverse{fb}, a luminosity which will certainly be surpassed at
the LHC.

\section{Conclusions}\label{Sec:conclusions}

We have simulated searches for SUSY at the LHC, assuming different
luminosities and centre of mass energies. We used a selection based
on the \MTTWO\ variable alone, an ATLAS MET-type search strategy and
a combined approach and applied them to $\sim$2000 pMSSM good-fit
points. Based on these we have suggested optimal values for the cut
on \MTTWO\ for low luminosities, which are 320~GeV for
$\sqrt{s}~=~7~\TeV$ and 400~GeV for $\sqrt{s}~=~14~\TeV$.

Ignoring systematic uncertainties in the background and using a
combined optimised \MTTWO\ and MET based search, we reach the
following conclusion. If we assign equal probability to each model
studied, a 7~TeV LHC would discover SUSY with 42.5\% probability
after accumulating 10~\inverse{fb}. Similarly, a 14~TeV LHC would
discover SUSY in 96.1\% of the cases after an integrated luminosity
of 10~\inverse{fb}. The few points that still evade detection are
characterised by heavy gluino masses above 2.5~TeV and heavy squark
masses or high mass degeneracies between the lightest squark and the
LSP.

The \MTTWO\ based strategy is better suited than the MET based
strategy for integrated luminosities up to several inverse
femtobarns, in the absence of systematic uncertainties in the
background. The \MTTWO\ strategy is based on a single cut on \MTTWO. It  accepts both more signal and more background than the MET strategy, meaning that at higher luminosities it is more vulnerable to fixed fractional systematic uncertainties in the background.
The MET based strategy is more sensitive than the \MTTWO\
based strategy to points with high gluino masses. Of course, a
combined strategy is the most powerful.

An optimization of the standard ATLAS searches (i.e. finding optimal values for the cuts on the selection variables used therein) would be desirable, though more difficult to make, due to the multi-dimensionality of the parameter space. A further optimization only makes sense in the context of good understanding of the systematic uncertainties, an important concern for the experiments.

While the LHC machine and the ATLAS and CMS experiments have performed remarkably well, SUSY searches at the LHC are still in their infancy. Our work
suggests that there is still much more of the parameter space to be
explored and that this can be done efficiently using the search
strategies available.

\section*{Acknowledgments}
We are grateful to the authors of \cite{AbdusSalam:2009qd} for providing us with the points analysed. Sasha Caron has kindly provided the significance calculator for section~\ref{Sec:systematics}. A.D. is thankful to Merton College, Oxford for hospitality. This work has been partially supported by STFC.

\bibliography{SUSYsearches}
\end{document}